# Femtosecond Quasiparticle and Phonon Dynamics in Superconducting $YBa_2Cu_3O_{7-\delta}$ Studied by Wideband Terahertz Spectroscopy

A. Pashkin,[1] M. Porer,[1] M. Beyer,[1] K. W. Kim,[1,2] A. Dubroka,[2] C. Bernhard,[2] X. Yao,[3] Y. Dagan,[4] R. Hackl,[5] A. Erb,[5] J. Demsar,[1,6] R. Huber,[1] and A. Leitenstorfer[1]

[1]*Department of Physics and Center for Applied Photonics, University of Konstanz, 78457 Konstanz, Germany*
[2]*Department of Physics, University of Fribourg, 1700 Fribourg, Switzerland*
[3]*Department of Physics, Shanghai Jiao Tong University, Shanghai 200240, China*
[4]*Raymond and Beverly Sackler School of Physics and Astronomy, Tel Aviv University, Tel Aviv, 69978, Israel*
[5]*Walther-Meissner-Institut, 85748 Garching, Germany*
[6]*Complex Matter Department, Josef Stefan Institute, Ljubljana, Slovenia*

We measure the anisotropic mid-infrared response of electrons and phonons in bulk $YBa_2Cu_3O_{7-\delta}$ after femtosecond photoexcitation. A line shape analysis of specific lattice modes reveals their transient occupation and coupling to the superconducting condensate. The apex oxygen vibration is strongly excited within 150 fs demonstrating that the lattice absorbs a major portion of the pump energy before the quasiparticles are thermalized. Our results attest to substantial electron-phonon scattering and introduce a powerful concept probing electron-lattice interactions in a variety of complex materials.

The interaction of electrons with the crystal lattice represents one of the most elusive, yet pivotal aspects of high-temperature superconductors (HTSCs). Although purely phonon-mediated BCS-type pairing fails to explain essential properties of superconducting (SC) cuprates, convincing evidence of significant electron-phonon contributions have been provided by angle-resolved photoemission [1–3], inelastic neutron scattering [4], tunneling [5] and Raman [6] spectroscopies. For time-integrated techniques it is difficult, however, to disentangle the interplay between elementary excitations.

Pump-probe studies have been harnessed to establish a temporal hierarchy of microscopic interaction processes in HTSCs following a strong optical perturbation. Near-infrared [7, 8], mid-infrared [9] and THz probe pulses [10] as well as Raman scattering [11] have been employed to study the recombination of photoexcited quasiparticles (QP) and the recovery of the SC condensate. A recent time-resolved photoemission study of $Bi_2Sr_2CaCu_2O_{8+\delta}$ has analyzed the relaxation of the quasi-equilibrium electronic temperature assuming a selective electron-phonon coupling [12]. However, since these experiments do not directly monitor the lattice degrees of freedom themselves, a more detailed picture of the role of the various phonon modes during an initial non-thermal regime has been beyond reach. Ultrafast electron diffraction has fuelled the hope to follow the evolution of the lattice directly [13], yet the time resolution has been limited to the picosecond scale so far.

Here we report a direct and simultaneous observation of QP and phonon resonances on the same femtosecond scale. Multi-THz few-cycle pulses monitor the mid-infrared conductivity of optimally doped $YBa_2Cu_3O_{7-\delta}$ (YBCO) as a function of the delay time $\tau$ after a near infrared excitation by 12 fs pulses. The resonant probe simultaneously traces the signatures of the SC gap, QP excitations, and two specific phonon modes. The dynamics of the phonon line shapes allows us to monitor vibrational occupations with a femtosecond resolution and to single out the coupling to the SC order parameter. We show that hot phonon effects and the creation of QPs both occur within 150 fs. This fact indicates extremely fast electron-phonon scattering in YBCO, in contrast to the usual assumption of a two-temperature model [14, 15].

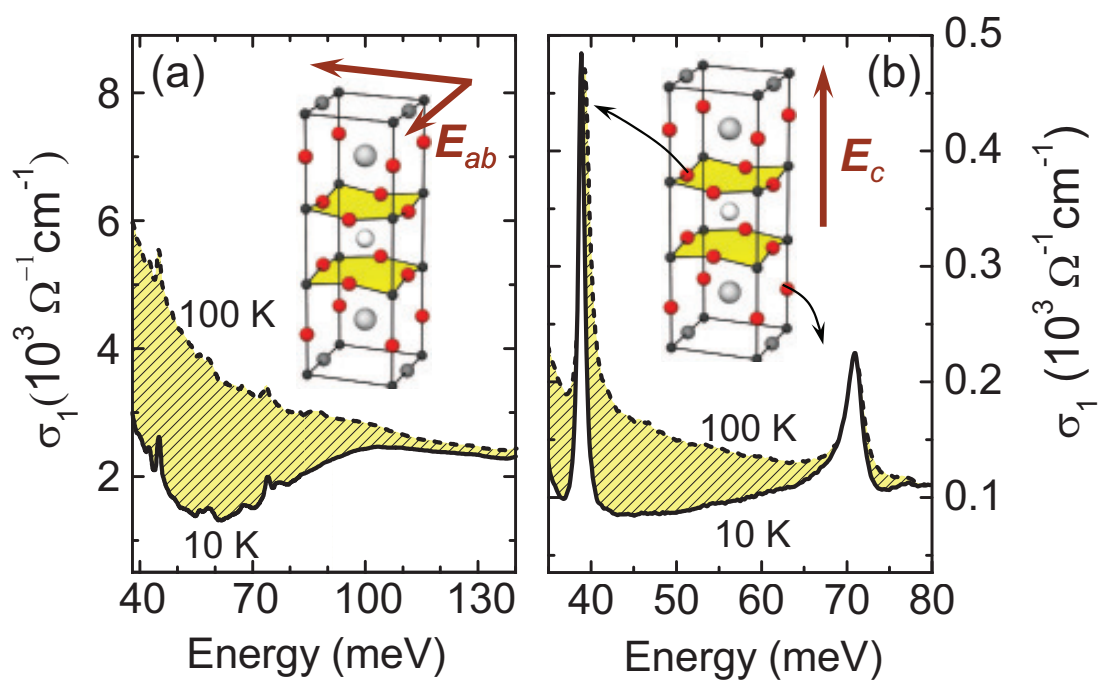

FIG. 1: Optical conductivity $\sigma_1(\omega)$ of YBCO in the SC ($T_L$ = 10 K, solid curve) and normal ($T_L$ = 100 K, broken curve) states for (a) $\mathbf{E}\perp c$ and (b) $\mathbf{E}\| c$. Insets: unit cell of YBCO and directions of the probe electric field. The oxygen ions involved in the observed vibrations are shown in red. The modes centered at 39 meV and 71 meV in (b) correspond to the bond-bending and apex oxygen vibrations, respectively.

Optimally doped single crystals of YBCO ($T_c$ = 92 K) form a prime laboratory for unconventional superconductivity. All measurements are carried out on as-grown surfaces oriented along or normal to the $c$ axis of the twinned system [16, 17]. As seen in the insets of Fig. 1, the di-rection of the probe field $\mathbf{E}$ may be chosen either parallel or perpendicular to $c$. Fig. 1(a) shows the real part of the in-plane conductivity ($\mathbf{E}\perp c$) in the normal and SC states, measured by ellipsometry [18]. As the temperature is decreased below $T_{\mathrm{c}}$, $\sigma_1$ drops sharply at energies $\hbar\omega < 110$ meV due to the opening of a SC gap [19]. A Drude-like contribution dominates the low-energy edge of the response even at $T_{\mathrm{L}} = 10\,K$, as expected for a $d$-wave order parameter where nodal QPs exist at finite temperatures. According to oscillator sum rules, the spectral weight defined by the highlighted area in Fig. 1(a) is proportional to the density of condensed QPs [19].

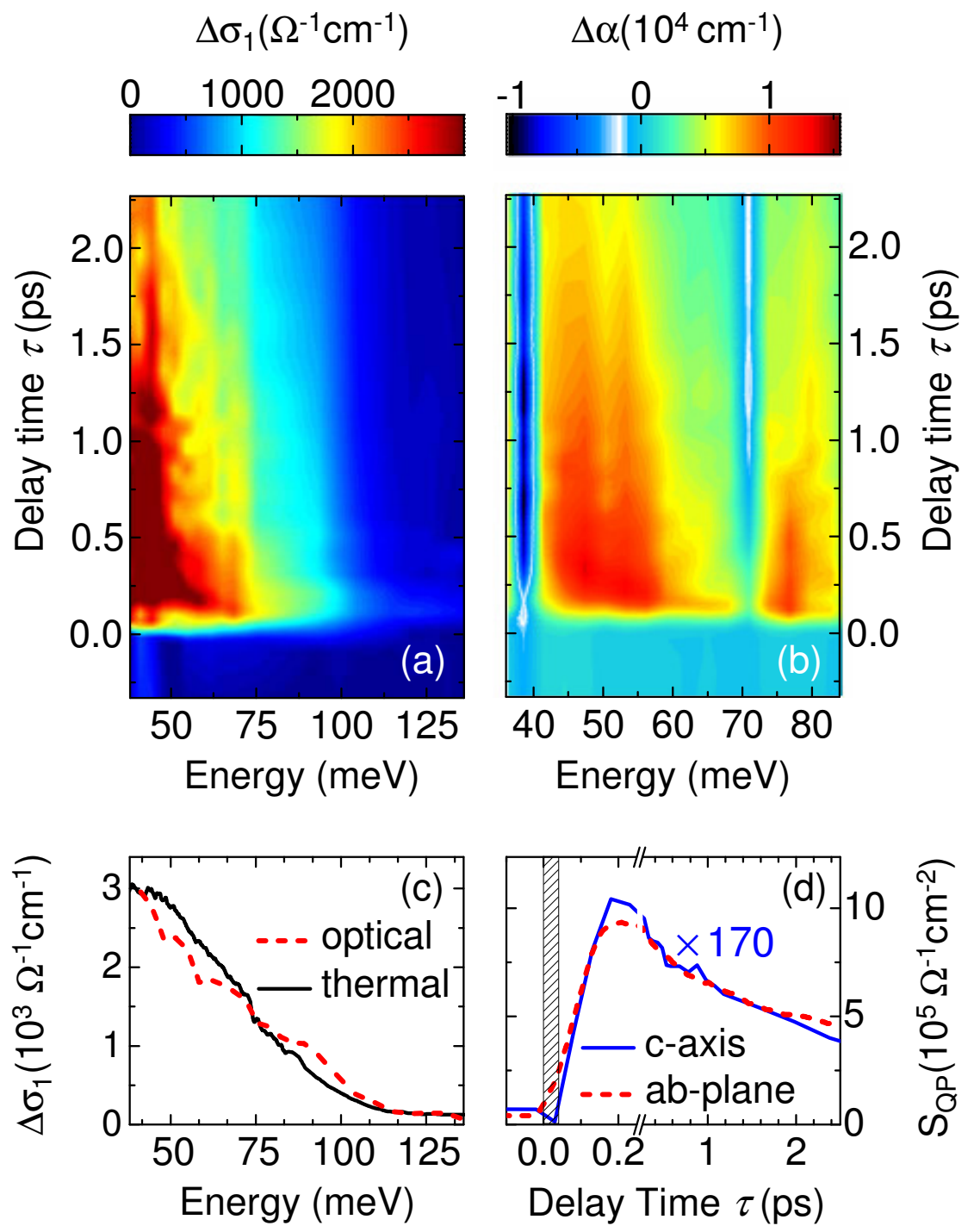

FIG. 2: 2D optical pump-THz probe data: (a) pump-induced changes $\Delta\sigma_1(\omega,\tau)$ as a function of the photon energy and the pump delay time $\tau$, for $\mathbf{E}\perp c$. (b) Corresponding changes of $\Delta\alpha(\omega,\tau)$, for $\mathbf{E}\| c$. Both experiments are performed at $T_{\mathrm{L}} = 20\,\mathrm{K}$ with $\Phi = 0.3\ \mathrm{mJ/cm^2}$. (c) Conductivity difference between normal and SC states (solid curve) for $\mathbf{E}\perp c$ and pump-induced $\Delta\sigma_1(\omega,\tau)$ at delay time $\tau = 1\,\mathrm{ps}$ after photoexcitation (broken curve). (d) Dynamics of the photoinduced QP spectral weight as a function the delay time $\tau$. Dashed curve: spectral weight for $\mathbf{E}\perp c$ integrated between 40 and 130 meV. Blue line: spectral weight (scaled by factor 170) for $\mathbf{E}\| c$ between 45 and 60 meV. The experimental time resolution is indicated by the hatched area.

While the in-plane response is dominated by electronic excitations, their contribution to the inter-plane optical conductivity ($\mathbf{E}\| c$) is reduced by one order of magnitude [see Fig. 1(b)]. This is why two phonon modes of $B_{1u}$ symmetry appear prominently on top of the electronic background: ($i$) The narrow resonance at $\hbar\omega = 39\,\mathrm{meV}$ is caused by Cu-O bond bending involving an in-phase motion of all oxygen ions within the cuprate bilayers. ($ii$) Collective vibrations of the apical oxygen ions located between the bilayers and the chains account for the conductivity maximum centered at $\hbar\omega = 71\,\mathrm{meV}$ [20]. The bond-bending mode displays an anomalous blue shift and a slight broadening [Fig. 1(b)] as the system is heated towards $T_{\mathrm{c}}$ [21–24]. At the same time the apex line changes from an asymmetric shape in the SC state to an almost symmetric one above $T_{\mathrm{c}}$ [22]. The origin of this effect will be discussed in more detail below.

In order to disentangle the ultrafast interplay between electrons and lattice, we perturb the SC state optically and trace the induced transition to the normal state with multi-THz transients. Details about our experimental setup are given in [17]. The unique stability of our laser system allows us to pursue, for the first time, field-sensitive reflection studies in bulk, single-crystalline YBCO of arbitrary orientation. The measured transient reflectivity provides direct access [25, 26] to the pump-induced changes of the complex conductivity spectra, $\Delta\tilde{\sigma}(\omega,\tau) = \Delta\sigma_1(\omega,\tau) + i\Delta\sigma_2(\omega,\tau)$, as a function of energy and the time delay $\tau$, without a Kramers-Kronig transform.

Figs. 2(a) and (b) depict the ultrafast response at a base temperature of $T_{\mathrm{L}} = 20\,\mathrm{K}$. The excitation fluence of $\Phi \simeq 0.3\ \mathrm{mJ/cm^2}$ is chosen to exceed the saturation threshold required to suppress superconductivity [17, 27]. We first discuss the transient in-plane response mapped out in panel (a). The pump pulse barely affects the high energy part of the conductivity whereas $\sigma_1(\omega)$ is enhanced below $\hbar\omega \approx 110$ meV due to the contribution of photocreated QPs. The absolute size and the spectral shape of the pump-induced signal corresponds to the conductivity difference between the SC and normal states [Fig. 2(c)]. The time evolution of the change of the spectral weight, $S_{\mathrm{QP}} = \int \Delta\sigma_1(\omega,\tau)d\omega$, between 40 and 130 meV is shown in Fig. 2(d). Interestingly, $S_{\mathrm{QP}}$ reaches its maximum with a delay of 150 fs, distinctly slower than our time resolution of 40 fs. These data present the first direct observation of the QP density generated during the pair-breaking cascade on the inherent time scale. The subsequent decay is described well by two exponential functions with time constants of 0.4 and 3.7 ps, respectively. The slow part corresponds to recondensation of QPs into Cooper pairs, whereas the fast component survives even above $T_{\mathrm{c}}$ and was associated to a pseudogap in Refs. [7, 9].

The situation for $\mathbf{E}\| c$ is yet richer in detail. While all essential features discussed in the following are also evident in $\Delta\sigma_1(\omega,\tau)$, the 2D map of the absorption coefficient $\Delta\alpha(\omega,\tau)$ illustrates the pump-probe dynamics most obviously [Fig. 2(b)]: Besides a broadband QP contribution featuring a similar spectral profile as in the case of $\mathbf{E}\perp c$, two narrow minima are superimposed at energies of 38 and 72 meV. These are the fingerprints of pump-induced anomalies of the phonon resonances. Remark-ably, the various contributions to the total response follow clearly different dynamics: The spectral weight of the QPs integrated in the energy range from 45 to 60 meV, i.e. between the phonon resonances, shows the same temporal trace as the in-plane conductivity [Fig. 2(d)]. In contrast, the pump-induced changes of the phonons are delayed. The most striking difference is imprinted on the apex phonon which exhibits its maximum anomaly as late as 1 ps after photoexcitation [Fig. 2(b)].

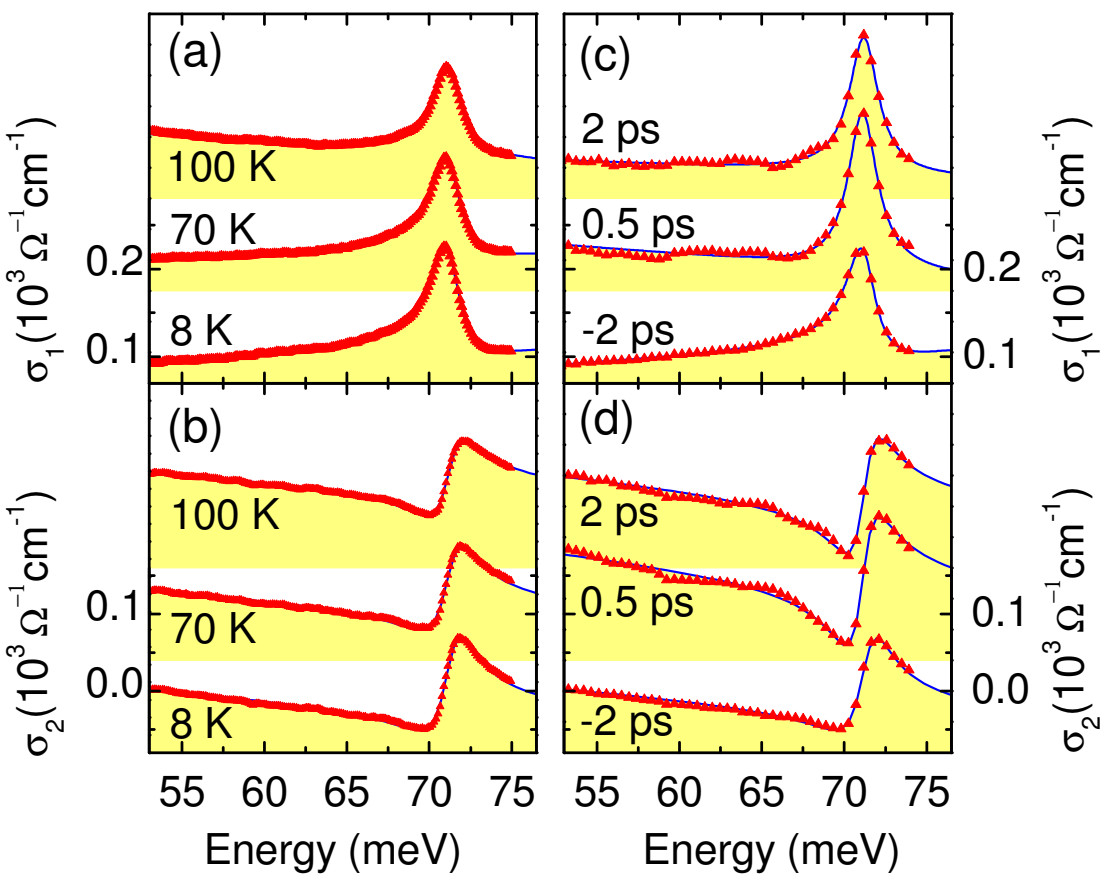

FIG. 3: Spectra of the optical conductivity $\sigma_1(\omega)$ and $\sigma_2(\omega)$ of YBCO for $\mathbf{E}\|c$: (a,b) equilibrium spectra at selected temperatures; (c,d) transient spectra at selected pump-probe delay times $\tau$ measured at $T_L = 20\,\mathrm{K}$ with $\Phi = 0.3\,\mathrm{mJ/cm^2}$. The solid lines show the fitting curves according to Eq. (1)

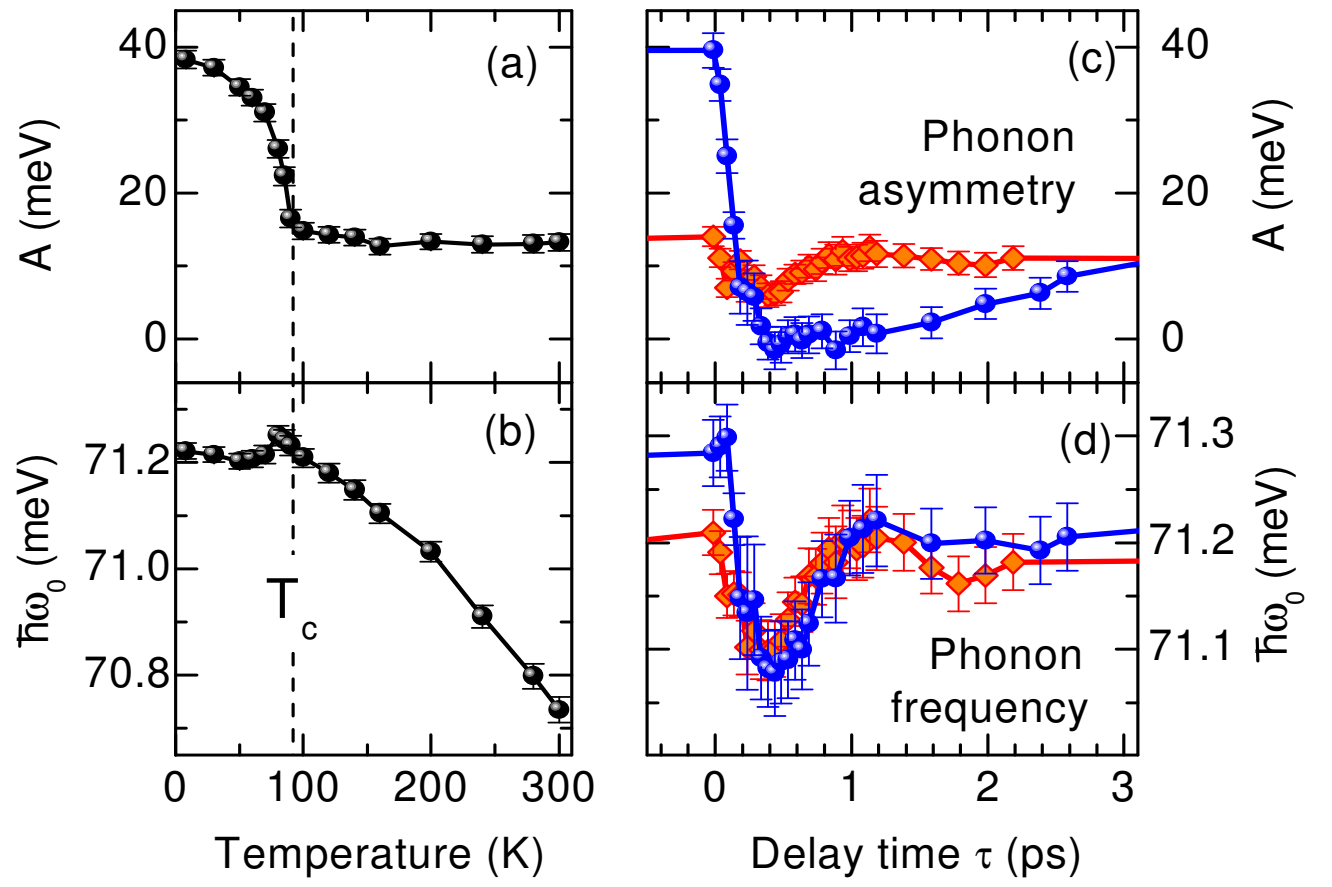

FIG. 4: (a) Asymmetry factor and (b) eigenfrequency of the apex mode as functions of temperature. (c) and (d) The corresponding quantities as functions of the pump-probe delay time $\tau$. The blue dots and red diamonds denote excitations of the SC ($T_L = 20\,\mathrm{K}$) and normal state ($T_L = 100\,\mathrm{K}$), respectively. The error bars indicate 95% confidence intervals for the fitting parameters.

Going beyond this qualitative observation, we introduce a line shape analysis of the apex mode which will allow us to single out the microscopic mechanisms underlying the pump-induced phonon anomaly. Fig. 3 depicts selected spectra of the real and imaginary part of the c-axis conductivity in the vicinity of the apex mode. Below $T_c$, $\sigma_1(\omega)$ features a characteristic asymmetric resonance. We describe the infrared response as the sum of a broadband electronic conductivity $\tilde{\sigma}_{el}(\omega)$ and a phenomenological model [28] of the phonon resonance:

$$\tilde{\sigma}(\omega) = \tilde{\sigma}_{el}(\omega) - \epsilon_0 S \frac{(\omega A + i\omega_0^2)\omega}{\omega_0^2 - \omega^2 - i\omega\gamma}, \qquad (1)$$

The phonon contribution is defined by the oscillator strength $S$, the eigenfrequency $\omega_0$, and the damping constant $\gamma$. The parameter $A$ accounts for the peculiar asymmetry of the lineshape below $T_c$. In order to reduce the number of free parameters, the QP background $\tilde{\sigma}_{el}(\omega)$ may be expressed by a linear function of frequency. Also, the damping $\gamma = 2\pi \times 0.5\,\mathrm{THz}$ is found to be almost temperature independent. The requirement to reproduce both real and imaginary parts of the spectra simultaneously imposes strict boundaries on the three remaining fit parameters $S$, $\omega_0$, and $A$. As seen in Figs. 3(a) and (b), the mid infrared response of YBCO in thermal equilibrium can be described convincingly with Eq. (1).

Figs. 4(a) and (b) display the resulting fit parameters $A$ and $\omega_0$ as a function of temperature. The asymmetry sharply decreases upon heating towards $T_c$ and levels off above this temperature. The physical origin of this phenomenon is explained by the coupling between optical phonons and the Josephson plasmon [24]. In the SC state the Josephson resonance renormalizes the phonon parameters by modifying local fields. In the case of the apex mode the dominant effect is a strong increase of the asymmetry parameter which, thus, quantifies the coupling of this mode to the Josephson plasmon and scales with the density of the SC condensate. The eigenfrequency $\omega_0$, on the other hand, shows a continuous red-shift with increasing temperature [Fig. 4(b)]. This feature is characteristic of an anharmonic lattice potential which softens for larger vibrational amplitudes. Consequently, the frequency shift of the apex phonon may be exploited as a sensitive probe of the vibrational occupation.

This idea opens up fascinating new perspectives in combination with ultrafast non-equilibrium measurements: Due to the excellent spectral resolution of our 2D multi-THz data, the model function of Eq. (1) can be numerically adapted to the transient conductivity spectra with high confidence as exemplified in Figs. 3(c,d). The dynamics of the phonon parameters is, thus, mapped out on a femtosecond scale [Figs. 4(c,d)]. Starting from the SC state [blue dots in Fig. 4(c)], the asymmetry of the apex mode decreases within $\tau \leq 150$ fs, mimicking the trace of condensate depletion. Analogously, the relaxation to the initial line shape asymmetry follows the QP recombination with a typical time constant of about 4 ps, confirming phonon-plasmon coupling as the microscopic origin of the asymmetry. Note that, unlike thermal activation above $T_c$, ultrafast optical excitation suppresses the asymmetry entirely for $\tau < 1$ ps. This fact represents a clear manifestation of the extremely non-equilibrium state of the system at early delay times.

Remarkably, the eigenfrequency of the apex mode [blue dots in Fig. 4(d)] experiences an abrupt red-shift within 150 fs after photoexcitation, i.e. during the onset of the electronic response, and reaches its extremum within 300 fs. The maximum photoinduced softening of the lattice resonance is comparable to the effect induced by a thermal phonon population at $T \approx 200$ K. We thus infer that the optical pump causes a hot phonon population of the apex mode. Subsequently, $\omega_0$ relaxes to a value which is slightly reduced with respect to the equilibrium level at $T_L = 20$ K. The relaxation time of 1 ps [see Fig. 4(d)] is substantially faster than the QP condensation, which further confirms that the phonon softening is not governed by the coupling to the Josephson plasmon. We suggest that the rapid recovery of $\omega_0$ is explained by a redistribution of the excess energy by phonon-phonon scattering into the large phase space of the entire Brillouin zone.

The damping of the apex mode corresponds to a phonon lifetime of 2 ps which exceeds the observed relaxation time significantly. This finding shows that phonon-phonon scattering is faster during the hot non-equilibrium state after photoexcitation as compared to a thermal distribution. Additional evidence for a hot phonon population as the main cause of the red-shift of the apex mode is provided by the dynamics in the normal state [red diamonds in Fig. 4(d)] where $\omega_0$ follows a quantitatively similar trace. The asymmetry parameter in the normal state, in contrast, shows a substantially reduced change owing to the absence of a macroscopic condensate [red diamonds in Fig. 4(c)] [29].

Our analysis, thus, allows us to disentangle the femtosecond dynamics of the QPs and specific lattice modes, for the first time. The results clearly indicate that a major portion of the absorbed pump energy is transferred to the phonon subsystem within 150 fs. Further evidence supporting this argumentation comes from an energy balance of the pump-induced transition into the normal state. The energy density of $\simeq 10$ J/cm$^3$ required to suppress the SC phase optically [17, 27] is much higher than the thermodynamically determined condensation energy of 1.2 J/cm$^3$ [30]. Following the lines of Ref. 8, only the phonon subsystem possesses sufficient heat capacity to dissipate the high excess energy of the pump pulse. It is worth comparing this situation with the so-called two- or three-temperature models [12, 15] which have been established assuming the electron-phonon scattering rate to be negligible as compared to electron-electron interaction. This hierarchy of scattering processes put forward in the context of normal metals [15] or HTSCs like $Bi_2Sr_2CaCu_2O_{8+\delta}$ [12] does not hold in case of YBCO where a hot phonon population and pair breaking occur on comparable time scales.

In conclusion, we present the first resonant femtosecond observation of both electronic and phononic degrees of freedom in a HTSC. Recording a photoinduced SC-to-normal transition with few-cycle multi-THz pulses provides qualitatively new insight into the electron-phonon interaction in YBCO. The lattice absorbs a large portion of the pump energy while the photoexcited charge carriers thermalize and the condensate is depleted. The results indicate strong electron-phonon scattering and introduce a powerful approach to measure the phonon occupation and the coupling to the SC condensate, on a sub-cycle temporal scale and in a broad variety of strongly correlated material systems.

We wish to thank W. Kaiser for inspiring discussions. This work has been supported by the Deutsche Forschungsgemeinschaft via the Emmy Noether Program and the Research Unit FOR538 (Er342/3 and Ha2071/3), as well as by the Alexander von Humboldt Foundation through a Sofja Kovalevskaja Award. The funding of the work at UniFr by SNF grants No. 200020-119784 and 200020-129484, at SJTU via SCST and the MOST of China (2006CB601003) and by the Kurt Lion foundation is gratefully acknowledged.

ever, the analysis is not as clear since the strong influence of the Josephson plasmon inhibits a complete separation of the anharmonic shift, see Supplementary Material.

# Supplementary material for "Femtosecond Quasiparticle and Phonon Dynamics in Superconducting $YBa_2Cu_3O_{7-\delta}$ Studied by Ultrabroadband Terahertz Spectroscopy"

A. Pashkin,[1] M. Porer,[1] M. Beyer,[1] K. W. Kim,[1,2] A. Dubroka,[2] C. Bernhard,[2] X. Yao,[3] Y. Dagan,[4] R. Hackl,[5] A. Erb,[5] J. Demsar,[1,6] R. Huber,[1] and A. Leitenstorfer[1]

[1]*Department of Physics and Center for Applied Photonics, University of Konstanz, 78457 Konstanz, Germany*
[2]*Department of Physics, University of Fribourg, 1700 Fribourg, Switzerland*
[3]*Department of Physics, Shanghai Jiao Tong University, Shanghai 200240, China*
[4]*Raymond and Beverly Sackler School of Physics and Astronomy, Tel Aviv University, Tel Aviv, 69978, Israel*
[5]*Walther-Meissner-Institut, 85748 Garching, Germany*
[6]*Complex Matter Department, Josef Stefan Institute, Ljubljana, Slovenia*

## SAMPLE PREPARATION

The single crystal of $YBa_2Cu_3O_{7-\delta}$ was grown by top-seeded solution growth [1]. The specimen were annealed, for 72 hours, in flowing oxygen at 500 °C and returned to room temperature by furnace cooling to obtain optimally oxygenated crystals. We performed all measurements on the specular as-grown surfaces of the crystal with an area of about $2 \times 2\,\mathrm{mm}^2$. One of these adjacent surfaces was oriented parallel, the other perpendicular, to the crystallographic $c$ axis of YBCO.

We also confirmed our key results by repeating the multi-THz experiments in an $a$-axis oriented, optimally doped $YBa_2Cu_3O_{7-\delta}$ film of a thickness of 500 nm. The sample was deposited on a 50 nm thick $PrBa_2Cu_3O_{7-\delta}$ template grown on a $LaSrGaO_4$ (100) substrate by RF sputtering. This process results in a highly oriented film in both the in-plane and out-of-plane directions [2]. For the latter only $a$-axis peaks were seen in x-ray diffractometry. We ensured that the in-plane $c$ and $b$ directions are well defined using x-ray diffraction at a grazing incidence. The critical temperature was determined by an AC susceptibility measurement as $T_c = 88 \pm 1\,\mathrm{K}$ .

## EXPERIMENTAL SETUP

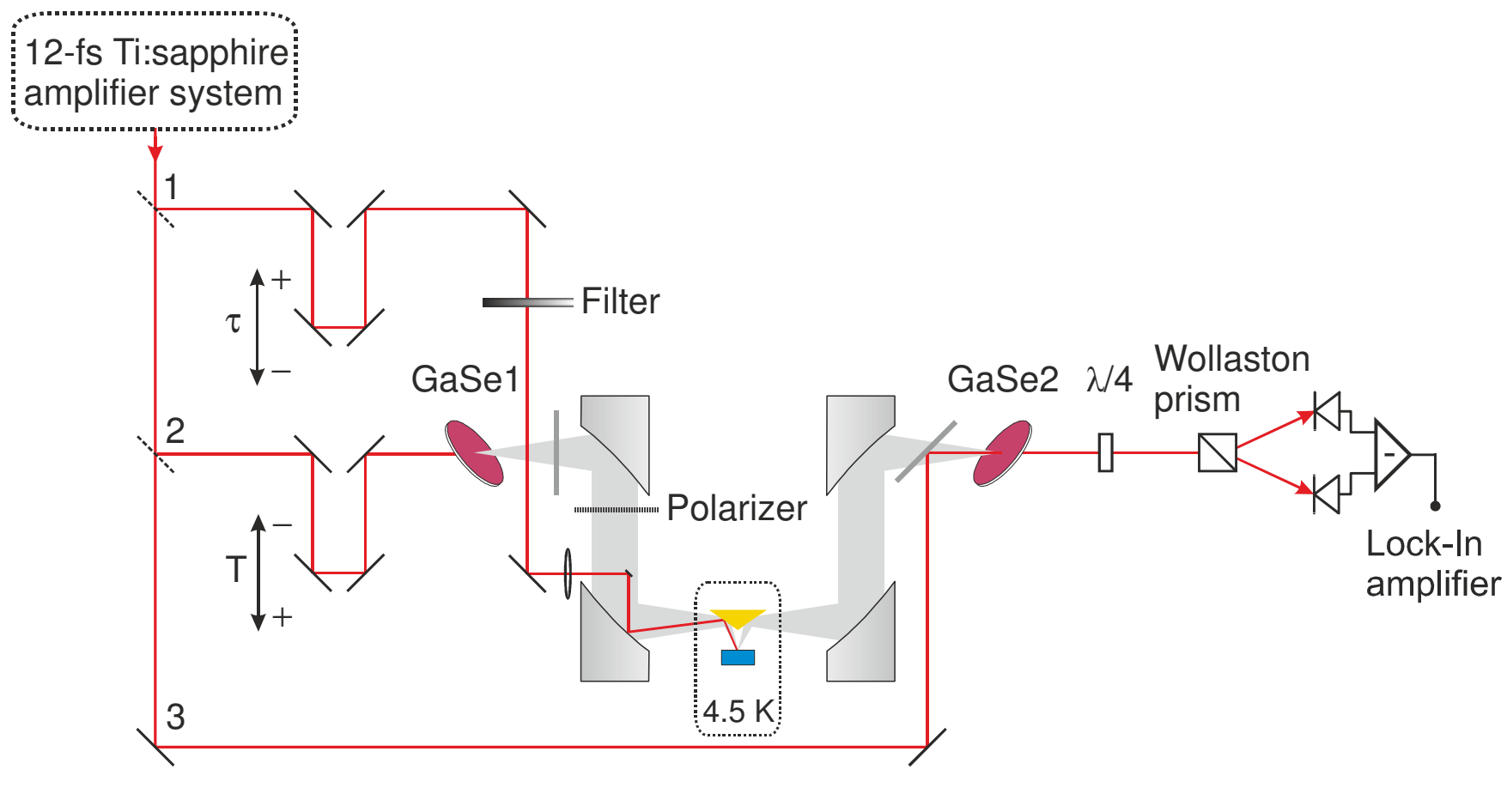

FIG. 1: Scheme of our NIR-pump / multi-THz-probe setup.

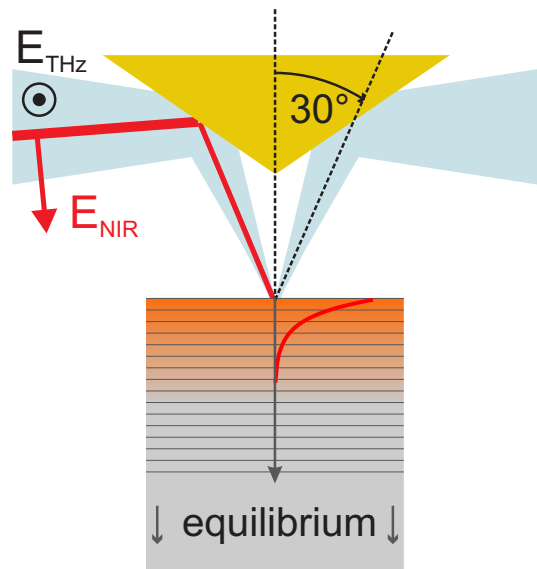

FIG. 2: Reflection geometry and model of the excitation profile.

Our setup is based on a home-built low-noise Ti:sapphire amplifier system for intense 12-fs light pulses centered at a photon energy of 1.55 eV [3]. The schematic of the NIR pump / multi-THz probe setup is depicted in Fig. 1. A first part of the laser output photoexcites the sample (branch 1). The diameter of the pump spot on the sample surface is set to 150$\mu$m (FWHM). The beam in branch 2 generates the probe transients via optical rectification in a 50 $\mu$m thick GaSe crystal. Off-axis parabolic mirrors focus the THz beam down to a spot size of 75 $\mu$m, which is two times smaller than the pump spot, ensuring homogeneous excitation of the probed surface area. Another two parabolic mirrors re-focus the reflected THz transient into the 50 $\mu$m thick GaSe electro-optic sensor. We trace the THz electric field amplitude by varying the delay $T$ of the transient with respect to the gating pulse (branch 3). For selected pump-probe delay times $\tau$ we simultaneously record a transient for both excited and unexcited sample. By dividing the Fourier spectra of these waveforms, we obtain two-dimensional maps of the pump-induced reflectivity change and the associated phase shift. The time resolution along the $\tau$-axis is limited by the detector bandwidth and amounts to 40 fs in our setup. The measurements are taken with $s$-polarized THz pulses and a $p$-polarized optical pump, both incident at an angle of 30° [see Fig. 2]. Thus, the pump is mostly polarized in the ab-plane of YBCO regardless of the orientation of the probe THz field ($\mathbf{E}\perp c$ or $\mathbf{E}\|c$).

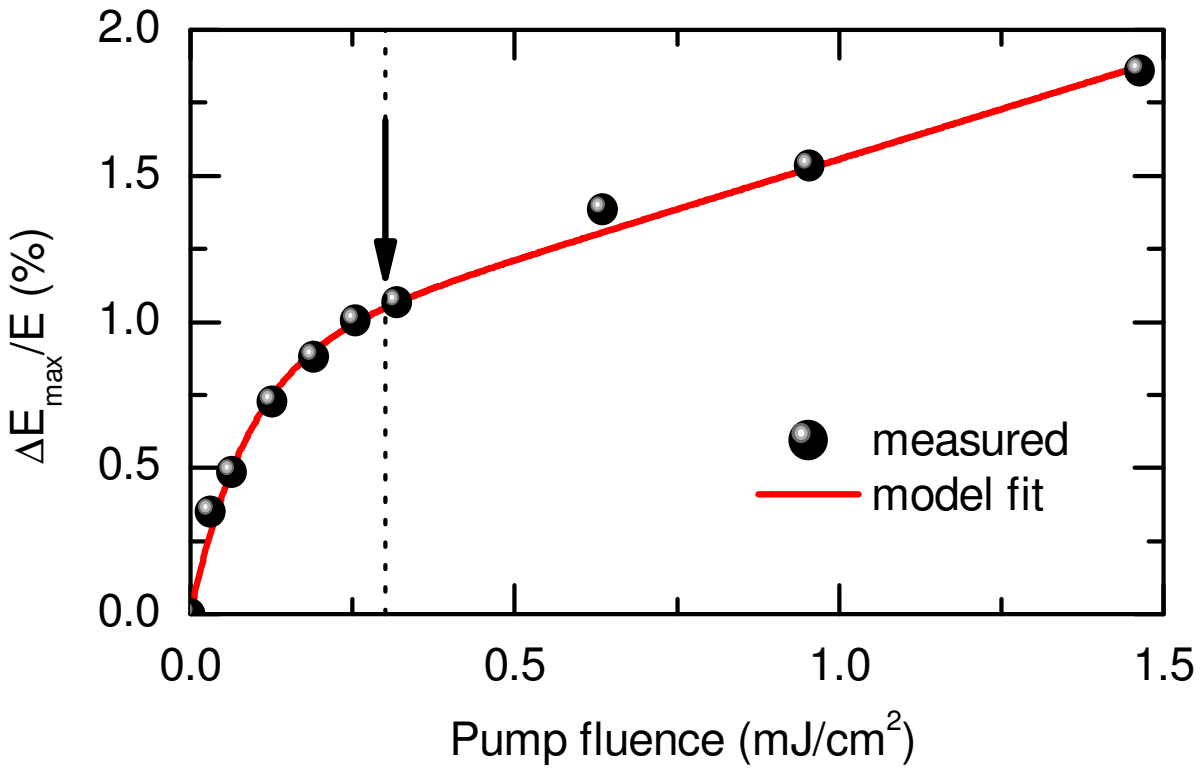

FIG. 3: Maximum pump-induced change of the reflected electric field for $\mathbf{E}\perp c$ as a function of the pump fluence. The arrow marks the pump fluence of $0.3\,\mathrm{mJ/cm^2}$ used for the experiment presented in the paper.

We employ a transfer-matrix formalism to extract the optical conductivity $\tilde{\sigma}(\omega,\tau)$ in the photoexcited state from the measured pump-induced changes in the complex reflectivity. An inhomogeneously excited surface layer is treated as a stack of much thinner layers with a homogeneous refractive index as shown by gray lines in Fig. 2. The excitation profile in $z$-direction is described by an exponential decay (red curve in Fig. 2). The refractive index is determined by numerical solution of the equation for the complex reflectivity. This method requires the knowledge of the equilibrium refractive index which is directly measured via ellipsometry.

## THE FLUENCE DEPENDENCE OF THE PUMP-PROBE RESPONSE

In order to determine the pump fluence which ensures full suppression of superconductivity and simultaneously keeps thermal heating of the sample low, we studied the fluence dependence of the maximal pump-induced change of the reflected THz-transients $\Delta E_{\mathrm{max}}(\Phi)$ prior to our two-dimensional scans. As one can see in Fig. 3, the pump-induced change shows a saturation behavior combined with a linear increase. This behavior can be well fitted by the following equation

$$\Delta E_{\mathrm{max}}(\Phi) = A\left(1 - \mathrm{e}^{-\Phi/\Phi_{\mathrm{sat}}}\right) + B_{\mathrm{lin}}\Phi. \tag{1}$$

The first term in Eq. 1 describes the depletion of the superconducting condensate characterized by the saturation fluence $\Phi_{\mathrm{sat}}$. The second linear term is related to the fast relaxation component of the quasiparticle response. Fitting the experimental data [Fig. 3] results in a saturation fluence of $\Phi_{\mathrm{sat}} = 0.1\,\mathrm{mJ/cm^2}$. The value of $\Phi = 0.3\,\mathrm{mJ/cm^2}$ chosen for the detailed analysis in the paper clearly exceeds the saturation threshold while it keeps the influence of Joule heating at a negligible level.

## EXCITATION DYNAMICS OF THE BOND-BENDING MODE

In contrast to the apex mode, the asymmetry of the bond-bending mode is small and only weakly affected by the Josephson plasmon. The dominant effect of plasmon coupling is given by a change of the eigenfrequency $\omega_0$. As temperature increases towards $T_{\mathrm{c}}$ the density of the SC condensate and its coupling to the bond-bending mode vanish leading to the blue-shift of $\omega_0$ as illustrated in Fig. 4(a). Pump-induced suppression of the SC state ($T_{\mathrm{L}} = 20$ K) also leads to a blue-shift of the bond-bending mode [Fig. 4(c)]. However, its dynamics does not reproduce the evolution of the QP response shown in Fig. 4(d) although the temperature dependence of $\omega_0$ below $T_\mathrm{c}$ clearly follows the QP spectral weight [compare Figs. 4(a) and 4(b)]. The initial rapid increase of the eigenfrequency takes about 300 fs, which is two times slower as compared to the depletion of the SC condensate. This stage is followed by a delayed growth of the eigenfrequency on the picosecond time scale. For $\tau > 2$ ps, the resonance frequency, finally, shifts back to its equilibrium value.

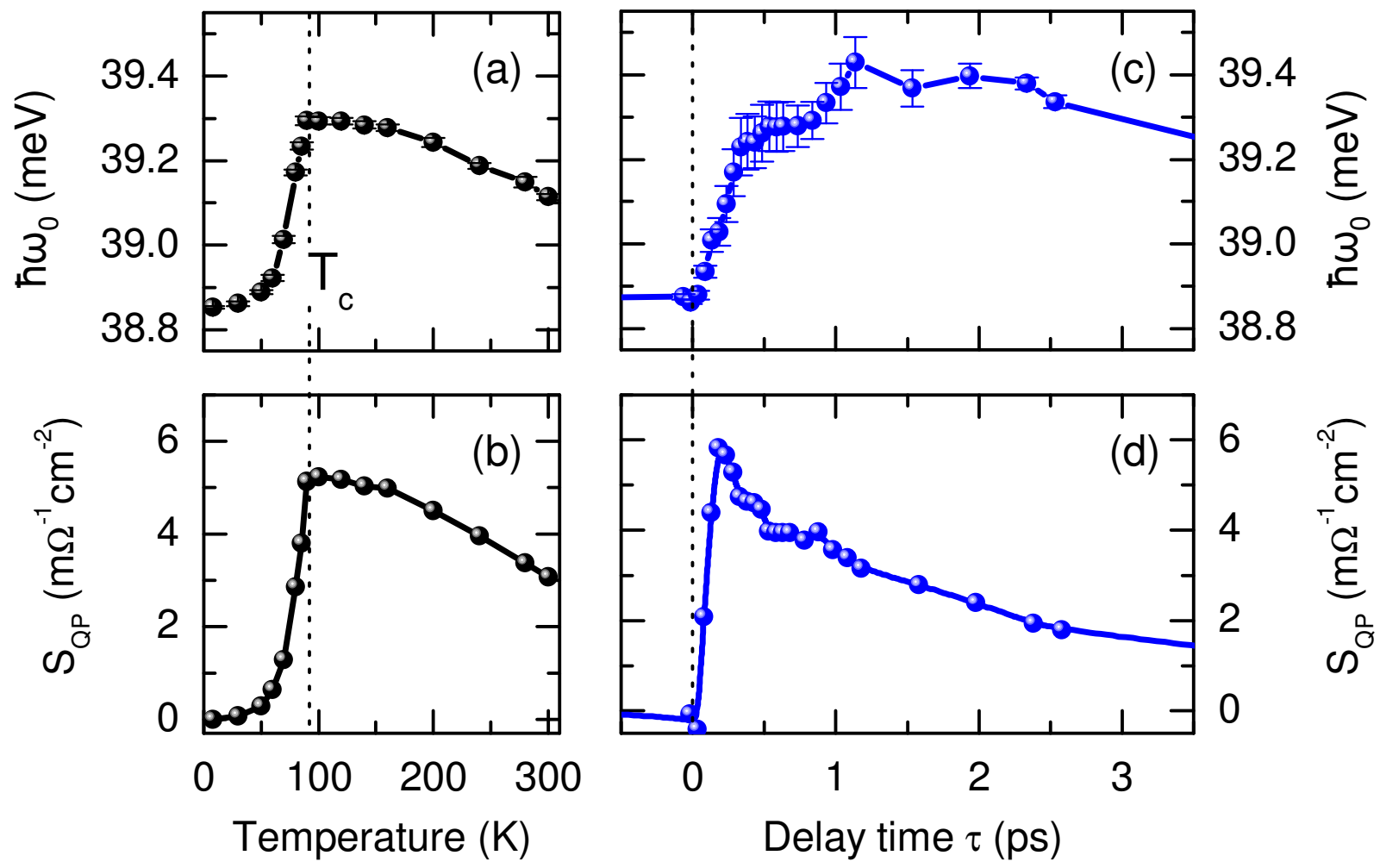

FIG. 4: Temperature dependencies of (a) the eigenfrequency of the bond-bending mode and (b) the QP spectral weight for $\mathbf{E}\|c$ integrated between 45 and 60 meV. (c) and (d) show the corresponding quantities as functions of the pump-probe delay time $\tau$ for the SC state ($T_\mathrm{L}$ = 20 K). The error bars indicate 95% confidence intervals for the bond-bending mode eigenfrequency.

Above $T_\mathrm{c}$, the bond-bending mode indicates a notable anharmonic red-shift [Fig. 4(a)], similar to the behavior of the apex phonon. This phenomenon is expected to persist also in the SC state. The effective blue-shift of the bond-bending resonance below $T_\mathrm{c}$, thus, results from the counteracting influence of Josephson coupling and lattice anharmonicities. The ultrafast dynamics depicted in Fig. 4(c) may be understood in these terms: Photoexcitation destroys the condensate and depletes plasmon coupling. The associated blue-shift of the phonon frequency is, therefore, expected to occur within 150 fs after the pump pulse. The delayed component of the dynamics of the frequency may be associated with the lattice anharmonicity induced by a hot phonon population. As the occupation number of the bond-bending mode decays, the red-shift relaxes and leads to a further increase of the phonon eigenfrequency, on a picosecond time scale. Finally, the resonance starts to return to its equilibrium frequency once the anharmonic shift becomes smaller than the effect of plasmon coupling. The relatively slow initial onset of the bond-bending mode anomaly within 300 fs evidences a different time of the electron-phonon interaction for this mode as compared to the apex mode.

Owing to the strong influence of the Josephson plasmon coupling on the frequency evolution of the bond-bending mode, it is not possible to disentangle it from the anharmonic population effect as clearly as in the case of the apex mode. Nevertheless, the distinct dynamics observed for the bond-bending and apex modes indicates a different interaction of these phonon modes with the non-equilibrium QPs. This fact allows us to exclude a scenario in which the phonon anomalies are caused by a photoinduced bond softening observed, for example, in bismuth [4, 5]. In this case, the onset dynamics for all phonon modes would be the same since it is governed by the density of the photoexcited electrons. Such a behavior is in contrast to our experimental observation of the phonon dynamics in YBCO.

## MEASUREMENT RESULTS FOR THE ORIENTED YBCO FILM

In order to verify the key findings for the QP and phonon dynamics we measured the pump-probe response of the $a$-axis oriented YBCO film. The experimental conditions are kept identical to the single crystal studies, including the excitation fluence of 0.3 mJ/cm$^2$ and the sample temperature of $T = 5$ K. The base temperature at the optically pumped surface of the sample could be kept as low as $T_\mathrm{L} = 5\,K$ due to the excellent thermal conductivity of the substrate.

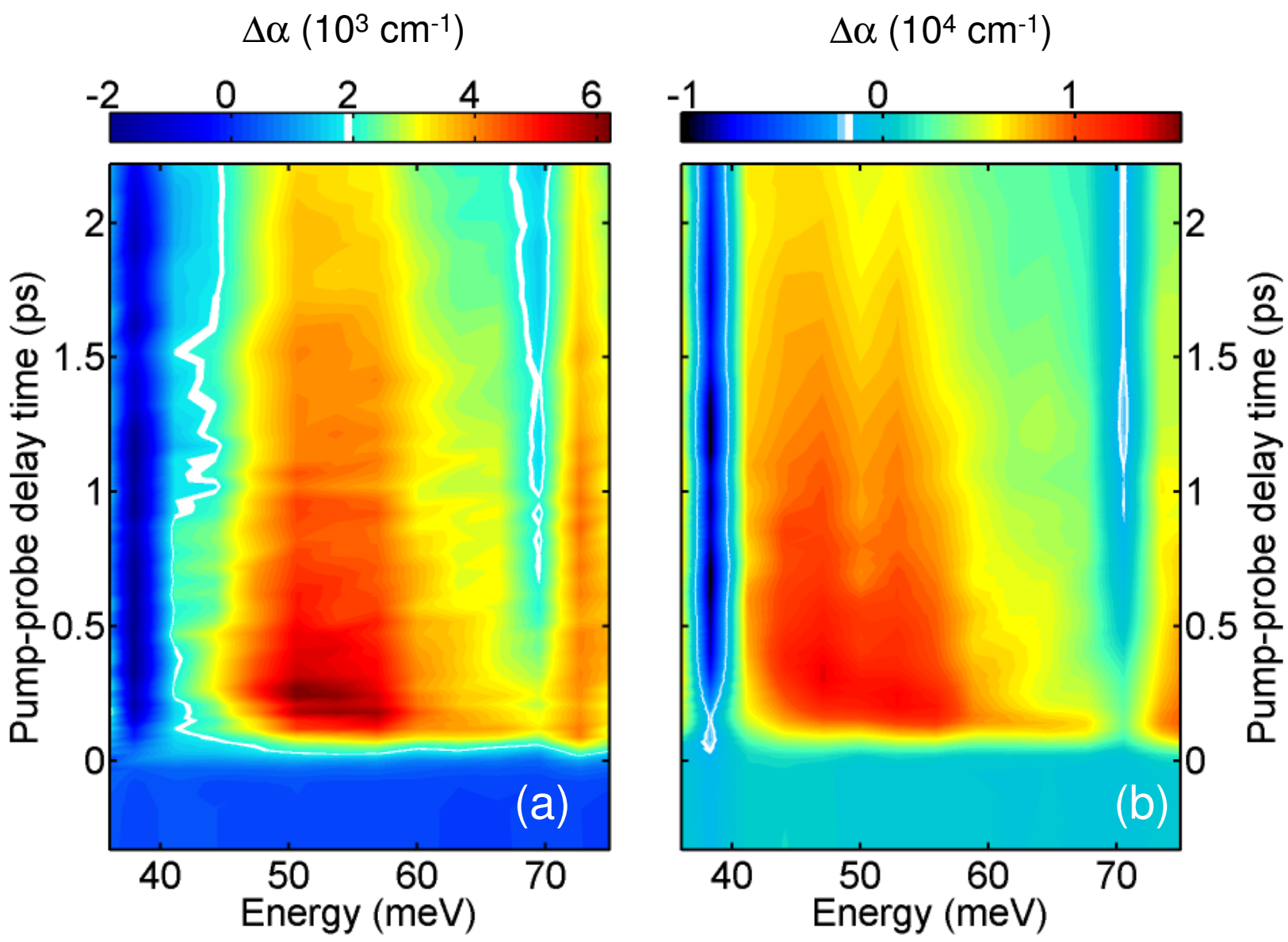

FIG. 5: 2D maps of the pump-induced changes of the absorption coefficient $\Delta\alpha(\omega, \tau)$ of (a) the $a$-oriented YBCO film and (b) the YBCO single crystal. Both experiments are performed with $\Phi = 0.3$ mJ/cm$^2$ and $\mathbf{E}\|c$. The base temperatures are kept far below $T_c$, at (a) $T_L = 5\,K$ and (b) $T_L = 20\,K$.

Fig. 5(a) depicts the pump-induced spectral change of the absorption coefficient $\Delta\alpha(\omega, \tau)$ of the YBCO film for $\mathbf{E}\|c$. The corresponding response of the single crystal is reproduced in Fig. 5(b) for comparison. Both 2D maps of $\Delta\alpha(\omega, \tau)$ demonstrate qualitatively the same behavior. The spectral response is characterized by a broadband QP contribution and two minima around the phonon energies of 38 and 69 meV. The spectral signatures of the phonon modes are weaker and broader as compared to the single crystal. This fact is caused by the broadening of the phonon lines due to phonon scattering on lattice imperfections of the film. Therefore, the results measured on the bulk crystal [Fig. 5(b)] have been chosen for the detailed analysis presented in the paper. Nevertheless, the time dynamics of the apex mode anomaly in the film is clearly delayed with respect to the QP response in the same way as in the YBCO crystal. This fact confirms the universality of the experimental results for $YBa_2Cu_3O_{7-\delta}$ presented in the paper.